\documentclass[reqno,11pt]{article}
\pdfoutput=1
\usepackage{jheppub}
\usepackage[T1]{fontenc}
\usepackage[latin9]{inputenc}
\usepackage{amsmath}
\usepackage{amssymb}
\usepackage{cancel}
\usepackage{stmaryrd}
\usepackage[dvipsnames]{xcolor}
\usepackage[normalem]{ulem}
\usepackage[tikz]{bclogo}
\usetikzlibrary{shapes, shadows, arrows, patterns, shadings} 
\usetikzlibrary{positioning}
\tikzset{mynode/.style={align=center,text width=3cm}}
\usepackage{pgf}
\usepackage{hyperref}
\usepackage{slashed}
\usepackage{shuffle}
\usepackage{babel}

\def\cd{\mathsf{cd}}

\title{One-loop double copy from gravity coupled to a massive vector field}
\author{Cristhiam Lopez-Arcos}
\affiliation{Institute of Physics of the Czech Academy of Sciences \& CEICO
~\\
Na Slovance 2, 18221 Prague, Czech Republic}
\emailAdd{cmlopeza@unal.edu.co}

\abstract{In this work, we compute the one-loop four-graviton amplitudes with a massive abelian vector field (Proca) circulating in the loop. Instead of the conventional Einstein-Hilbert formulation, we employ the Landau-Lifshitz metric density approach, coupling gravity to the Proca field. Within this framework, we found that the resulting contact-terms-free $n$-gon numerators exhibit a manifest double-copy structure, provided the external gravitational trees satisfy this property. This structure enables the direct construction of the corresponding amplitudes with a scalar in the loop. We further show that these numerators can be consistently used to evaluate pure gravity amplitudes in the sectors where contact terms vanish, offering a simplified setting to explore double-copy relations at one loop level.}

\begin{document}

\maketitle

\tableofcontents{} 

\section{Introduction}

The perturbative formulation of a quantum field theory for gravity presents a major technical challenge from the outset: it involves an infinite number of interaction vertices, which leads to a rapid proliferation of Feynman diagrams \cite{Feynman:1996kb}. Moreover, for many years progress was also hindered by another fundamental issue, the theory's lack of renormalizability \cite{tHooft:1974toh}.

Loop amplitudes help us to understand the ultraviolet (UV) behaviour of gravity. One-loop calculations can reveal cancellations and potential underlying structures that hint at a consistent UV completion. Additionally, such amplitudes also contribute to the post-Minkowskian expansion of general relativity, which is crucial for modelling gravitational wave emissions from binary systems. Recent results have shown a deep connection between loop-level amplitudes and classical observables \cite{Brandhuber:2023hhy}. Going to higher loop cases, the two-loop propagator contributes to the low-energy effective action of gravity, influencing quantum corrections to classical observables such as black hole interactions and gravitational wave signals \cite{Donoghue:2022eay}.

In this context, the study of gravitational loop amplitudes with different kinds of particles, massive or massless, running in the loop becomes particularly relevant. Each type of field distinctively contributes to the ultraviolet and infrared behaviour of gravity, providing valuable information about its quantum consistency and possible effective descriptions. Additionally, these mixed matter-gravity loops allow us to probe how quantum fields of various spins couple to the gravitational field, offering insights into both the universality of gravitational interactions and their implications for phenomenological models, including those involving massive vector or scalar fields from dark sector extensions (see \cite{Jaeckel:2016jlh} and references therein). Proca fields in particular have also found applications in astrophysics \cite{Brito:2015pxa}.

Another particular aspect of a quantum theory of gravity is that it enjoys the possibility of field redefinitions \cite{Ananth:2007zy,Knorr:2023usb}. One of these redefinitions is the tensor density one proposed by Landau and Lifshitz \cite{Landau:1975pou}, which has found applications in post-Newtonian calculations \cite{Blanchet:2013haa}. In the perturbative formulation of gravity, the tensor density formulation has made few appearances. It was used in the calculation of the graviton self-energy in pure gravity and coupled to a Maxwell field by Capper et al. \cite{Capper:1973pv,Capper:1974ed}. Recently, it also found an application in the recursive formulation for one-loop gravity integrands \cite{Gomez:2024xec}.  

An important reason for the limited use of the tensor density formulation in perturbative field theory is the difficulty of coupling it to fermionic fields, which is not as straightforward as in the metric formulation. This complication prevents a direct reformulation of supergravity theories within this framework. Moreover, the metric coupling with the veilbein arises naturally in string theory. The reason to bring strings into the discussion is because in its context a new paradigm for gravitational amplitudes originated: The double copy, which allow us to understand gravity as the \emph{square} of gauge theory \cite{Kawai:1985xq}. This double-copy construction has since found a more refined realization in field theory through colour-kinematics duality, as shown by Bern, Carrasco and Johansson (BCJ) \cite{Bern:2008qj,Bern:2010ue}.

The double-copy construction based on colour-kinematics duality has provided a powerful tool to relate gauge and gravity amplitudes at the integrand level. While tree-level amplitudes can be efficiently expressed in terms of kinematic numerators that satisfy Jacobi-like relations mirroring those of the colour factors, the situation becomes considerably more intricate at one loop. In particular, for non-supersymmetric theories, constructing one-loop master numerators that satisfy colour-kinematics duality has long been a challenging open problem, remaining unresolved beyond the four-point case \cite{Bern:2013yya,Nohle:2013bfa} for over a decade, until the recent breakthrough at five points presented in \cite{Cao:2025ygu}. More progress in this direction has been achieved mainly in highly symmetric settings, such as maximally supersymmetric Yang-Mills and supergravity theories \cite{Edison:2022jln}. A practical way to circumvent the explicit construction one-loop BCJ master numerators (the ones that satisfy the same algebraic relations of their respective colour factor) is through the unitarity method \cite{Bern:1994zx,Bern:1994cg}, where loop amplitudes are reconstructed from their on-shell cuts, effectively encoding the double-copy structure without requiring a manifest one-loop dual representation. The computation of four-graviton amplitudes with massive particles of spin up to 2 circulating in the loop using unitarity methods was presented in \cite{Bern:2021ppb}, although in that work the vector field in the loop corresponds to a non-abelian case.

The Landau-Lifshitz formulation offers notable advantages when coupled to bosonic fields, as part of the metric-dependent measure is absorbed into the definition of the metric density, leading to simpler vertices. For instance, it provides one of the simplest gravitational interactions when coupled to a massless scalar field. Moreover, in this formulation the de Donder gauge condition takes a Lorenz-like form, yielding further simplifications, a feature we exploit here by coupling the metric density to an abelian massive vector field (Proca field \cite{Proca:1936fbw}). In this work, we further circumvent the difficulty of constructing full one-loop numerators satisfying colour-kinematics duality by imposing this condition only on the external trees attached to the loop, while keeping the loop numerators in a simpler abelian form. This strategy preserves the advantages of the double-copy structure while avoiding the complexity of constructing a full one-loop in the representation that satisfies colour-kinematics duality.

Finally, the integrations will be carried out using the traditional Passarino-Veltman (PV) \cite{Passarino:1978jh} tensor reduction method, which is very well implemented in Mathematica packages like Package-X \cite{Patel:2016fam} and FEYNCALC \cite{Mertig:1990an}. The notable simplicity of the integrands obtained through our approach eliminates the need for more sophisticated unitarity-based techniques, which are often required in loop calculations. This allows for a direct and transparent evaluation of the amplitudes while preserving the underlying one-loop double-copy structure.

This work is organized as follows: On the second section we will present the methods and conventions that will be used in our calculations. In the third section we will present the action for the Proca field coupled to gravity and will construct the one-loop integrands from its equations of motion. The fourth section will be a brief review of tree-level colour-kinematics and double copy. In the fifth section we will apply the tree-level double copy on the one-loop numerators to study the double copy that appear on these, inherited from the external trees. Section six will be devoted to the calculation of four-graviton amplitudes with scalar and vector particles in the loop from the numerators found in the previous section. The seventh section will be an exploration to one-loop Pure Yang-Mills and pure gravity amplitudes from single and double copy. In section eight we conclude.

\section{Calculation generalities}

In this section, we outline the computational tools employed to evaluate the amplitudes presented in this work. The analysis is based exclusively on traditional techniques of integration, since our primary objective is the explicit computation of one-loop four-point amplitudes.

The integration method we will use is the PV tensor reduction \cite{Passarino:1978jh}. The idea is to take one-loop rank-$r$ tensor integrals in $D=4-2\epsilon$ dimension, with $n$ propagators $\Delta_i$ and $k_i$ the external particles momenta
\begin{equation}\label{eq:tensor-int}
I_n^{\mu_1\cdots\mu_r} =
\int \frac{d^D\ell}{i\pi^{D/2}} \,
\frac{\ell^{\mu_1}\cdots \ell^{\mu_r}}
{\Delta_0 \Delta_1 \cdots \Delta_{n-1}},
\qquad
\Delta_i = (\ell + k_i)^2 - m_i^2,
\end{equation}
then reduce them to a linear combination of one-loop scalar integrals, denoted $A_0$ (tadpole), $B_0$ (bubble), $C_0$ (triangle), and $D_0$ (box).

The result of the reduction procedure leads to an expression for the tensor integral \eqref{eq:tensor-int} as
\begin{equation}\label{eq:pv-red}
I_n^{\mu_1\cdots\mu_r} =
\sum_{i_1,\dots,i_r} 
T^{\mu_1\cdots\mu_r}F_{i_1,\dots,i_r},
\end{equation}
where $T^{\mu_1\cdots\mu_r}$ are all the possible symmetric tensors built from metric $\eta^{\mu\nu}$ and external momenta $k_i^\mu$ and the $F_{i_1,\dots,i_r}$ are the coefficient functions in terms of the Mandelstam variables. For such variables we will adapt the following conventions
\begin{eqnarray}
(k_1+k_2)^2 &=& s,\\
(k_2+k_3)^2 &=& t,\\
(k_1+k_3)^2 &=& u,
\end{eqnarray} 					
which due to momentum conservation satisfy $s+t+u=0$.

For the simplifications in the kinematics we will not use the spinor-helicity technique \cite{Berends:1981rb} (see also \cite{Dixon:1996wi}) explicitly, but a set of cancellation rules in the kinematic invariants obtained from it after a suitable choice of reference momenta for the polarizations. For the three configurations of the four-point amplitude we will use the following sets of rules:
\begin{itemize}
\item[1)] For the all-plus $(++++)$ case we take the all the polarizations with the same reference momentum $q^{\mu}=-2tk^{\mu}_1+(s-t)(2k^{\mu}_2+k^{\mu}_3)$ and obtain the following set of cancellation rules
\begin{eqnarray}
&&\varepsilon_i\cdot\varepsilon_j=0,\,\varepsilon _1\cdot k_2= -\frac{1}{2}\varepsilon _1\cdot k_3,\,\varepsilon _1\cdot k_4= -\frac{1}{2} \varepsilon_1\cdot k_3,\notag\\
&&\varepsilon _2\cdot k_1= \frac{\left(k_1\cdot k_2-k_2\cdot k_3\right) \varepsilon _2\cdot k_3}{2 k_2\cdot k_3},\,\varepsilon _2\cdot k_4=-\frac{\left(k_1\cdot k_2+k_2\cdot k_3\right) \varepsilon _2\cdot k_3}{2 k_2\cdot k_3},\notag\\
&&\varepsilon _3\cdot k_1=\frac{\left(k_1\cdot k_2-k_2\cdot k_3\right) \varepsilon _3\cdot k_2}{k_2\cdot k_3},\,\varepsilon _3\cdot k_4= -\frac{k_1\cdot k_2 \varepsilon_3\cdot k_2}{k_2\cdot k_3},\notag\\
&&\varepsilon _4\cdot k_2= -\frac{\left(k_1\cdot k_2+k_2\cdot k_3\right) \varepsilon _4\cdot k_3}{2k_1\cdot k_2},\,\varepsilon _4\cdot k_1= \frac{\left(k_2\cdot k_3-k_1\cdot k_2\right) \varepsilon _4\cdot k_3}{2 k_1\cdot k_2}.
\end{eqnarray}
\item[2)] For the mostly-plus $(-+++)$ case we take the reference momenta as $(k_4,k_1,k_1,k_1)$ and obtain
\begin{eqnarray}
&&\varepsilon_i\cdot\varepsilon_j=0,\,\varepsilon_1\cdot k_4= 0,\,\varepsilon_2\cdot k_1= 0,\,\varepsilon_3\cdot k_1= 0,\,\varepsilon_4\cdot k_1= 0,\notag\\
&&\varepsilon_1\cdot k_3=-\varepsilon_1\cdot k_2,\,\varepsilon_2\cdot k_4=-\varepsilon_2\cdot k_3,\varepsilon_3\cdot k_4= -\varepsilon_3\cdot k_2,\varepsilon_4\cdot k_3=-\varepsilon_4\cdot k_2
\end{eqnarray}
\item[3)] For the Maximum Helicity Violating (MHV) $(--++)$ case the reference momenta will be $(k_4,k_4,k_1,k_1)$, this implies
\begin{eqnarray}
&&\varepsilon_1\cdot \varepsilon_2= 0,\,\varepsilon_1\cdot \varepsilon_3= 0,\,\varepsilon_1\cdot \varepsilon_4= 0,\,\varepsilon_2\cdot \varepsilon_4= 0,\varepsilon_3\cdot \varepsilon_4= 0,\,\varepsilon_2\cdot \varepsilon_3= \frac{\varepsilon_2\cdot k_3 \epsilon_3\cdot k_2}{k_2\cdot k_3},\notag\\
&&\varepsilon_1\cdot k_4= 0,\,\varepsilon_2\cdot k_4= 0,\,\varepsilon_3\cdot k_1= 0,\,\varepsilon_4\cdot k_1=0,\,\varepsilon_1\cdot k_3= -\varepsilon_1\cdot k_2,\notag\\
&&\varepsilon_2\cdot k_3= -\varepsilon_2\cdot k_1,\,\varepsilon_3\cdot k_4= -\epsilon_3\cdot k_2,\,\varepsilon_4\cdot k_3= -\varepsilon_4\cdot k_2
\end{eqnarray}
\end{itemize}

With this we summarize the techniques for the calculation of the amplitudes and also the conventions that we will use in this work.

\section{Gravity coupled to a Proca field}

Before coupling gravity to a Proca field we review the gravitational action with the change of variables to the Landau-Lifshitz prescription. For the starting point we take the original formulation of general relativity with the Einstein-Hilbert action, where we have the metric $g^{\mu\nu}$ as the fundamental variable (the gravitational field). We cast the original gravitational action as
\begin{equation}
S_{\text{EH}}=\int d^{4}x\,\sqrt{-g}R,\label{eq:EH-action}
\end{equation}
where, $g$ is the determinant of the inverse metric $g_{\mu\nu}$, and $R=g^{\mu\nu}R_{\mu\nu}$
is the scalar curvature. The Ricci tensor is the trace of the Riemann tensor $R_{\mu\nu}=R_{\mu\rho\nu}^{\rho}$,
with the latter given by
\begin{equation}
R_{\mu\rho\nu}^{\sigma} = \partial_{\rho}\Gamma_{\mu\nu}^{\sigma}-\partial_{\nu}\Gamma_{\rho\mu}^{\sigma}+\Gamma_{\rho\gamma}^{\sigma}\Gamma_{\mu\nu}^{\gamma}-\Gamma_{\nu\gamma}^{\sigma}\Gamma_{\mu\rho}^{\gamma},
\end{equation}
and $\Gamma^{\rho}_{\mu\nu} = \frac{1}{2}g^{\rho \sigma}(\partial_{\mu}g_{\nu\sigma}+\partial_{\nu}g_{\mu\sigma}-\partial_{\sigma}g_{\mu\nu})$.

The idea of Landau-Lifshitz \cite{Landau:1975pou} was to write the energy-momentum tensor for the gravitational field as a tensor density or pseudotensor, as they called it. At the level of the gravitational field equations, they introduced the metric tensor density
\begin{equation}\label{eq:gothicg}
\mathfrak{g}^{\mu\nu}=\sqrt{-g}g^{\mu\nu},
\end{equation}
also known as \emph{gothic g}, and then interpreted it as the fundamental field. 

An Action for the metric density \eqref{eq:gothicg} was found later by Goldberg \cite{Goldberg:1958zz}. We will see that such action seems more suitable for perturbation theory, since we would not have to deal with the long expression for the interaction vertices that come from expanding the measure of integration in \eqref{eq:EH-action}. Replacing \eqref{eq:gothicg} into \eqref{eq:EH-action} we obtain the shorter version of the gravitational action as we can see now 
\begin{eqnarray}\label{eq:LL-action}
S_{\text{LL}}=\int d^{4}x \, \mathfrak{g}^{\mu\nu}\big(\partial_{\mu}\mathfrak{g}^{\rho\sigma}\partial_{\nu}\mathfrak{g}_{\rho\sigma}-2\partial_{\mu}\mathfrak{g}^{\rho\sigma}\partial_{\rho}\mathfrak{g}_{\nu\sigma}+\frac{1}{2}\mathfrak{g}^{\rho\sigma}\partial_{\mu}\mathfrak{g}_{\rho\sigma}\mathfrak{g}^{\gamma\lambda}\partial_{\nu}\mathfrak{g}_{\gamma\lambda}\big). 
\end{eqnarray}
Another advantage of this field redefinition can be seen in the de Donder gauge condition. In the original formulation we have the non-linear relation $g^{\mu\rho}\Gamma^{\nu}_{\mu\rho}=0$, while in the metric density approach acquires the simpler form
\begin{eqnarray}
\partial_{\mu}\mathfrak{g}^{\mu\nu}=0,
\end{eqnarray}
that remind us of the Lorenz condition for gauge fields.

We can easily calculate the tree-level amplitudes, or the trees that will be attached to the loop, from the equations of motion from  
\eqref{eq:LL-action}. Since those trees will not be used here, we do not have to go into much detail and just briefly explain how to obtain them (those can be found, even completely off-shell in \cite{Gomez:2024xec} and for the original gravitational action in \cite{Gomez:2021shh}). The idea, following the procedure from \cite{Gomez:2021shh}, is to calculate the equation of motion from \eqref{eq:LL-action} and then replace the \emph{multi-particle} formal solutions 
\begin{eqnarray}\label{eq:forsol-g}
\mathfrak{g}^{\mu\nu}(x) & = & \eta^{\mu\nu}-\sum_{P}I_{P}^{\mu\nu}e^{ik_{P}\cdot x},\\
\mathfrak{g}_{\mu\nu}(x) & = & \eta_{\mu\nu}+\sum_{P}H_{P\mu\nu}e^{ik_{P}\cdot x},
\end{eqnarray}
where $P=i_1i_2\dots i_{|P|}$ is now a multi-index, also known as a word, of size $|P|$ and $k_{P}=k_{i_1}+k_{i_1}+\dots+k_{i_{|P|}}$. The single particle case $I_k^{\mu\nu}=h_k^{\mu\nu}$ is what we identify as the quanta of the gravitational field, i.e. the graviton. With this choice, for $H_{k\mu\nu}=h_{k\mu\nu}$ we have a geometric series in $h_k^{\mu\nu}$. From the recursive solution of the equations of motion, the multi-particle $I_{P}^{\mu\nu}$ fields obtained are the gravitational version of the Berends-Giele currents, that were originally found for Yang-Mills in \cite{Berends:1987me}. These currents now will enter as the external trees that will be attached to the loop. From such trees we will only use the important condition that comes from imposing the de Donder gauge on the currents, i.e. 
\begin{eqnarray}
k_{P\mu}I_{P}^{\mu\nu}=0.
\end{eqnarray}

Using the definition \eqref{eq:gothicg} we can also find a relation between the currents from both gravity actions. If we denote the currents from the Einstein-Hilbert action \eqref{eq:EH-action} by $\tilde{I}_P^{\mu\nu}$, these will be related to $I_P^{\mu\nu}$ by
\begin{eqnarray}\label{eq:LL-EH-curr}
I_P^{\mu\nu}=\tilde{I}_P^{\mu\nu}-\tilde{E}_P\eta^{\mu\nu}+\sum_{P=Q\cup R}\tilde{E}_Q\tilde{I}_R^{\mu\nu}
\end{eqnarray}
where the scalar factors $\tilde{E}_P$ come from the expansion of the $\sqrt{-g}$, that we will explain later, and the sum runs over the decomposition of the word $P$ into the ordered words $\{Q,R\}$. In general, we can decompose $P$ into any number of ordered words $\{Q,R,S\dots\}$ as $P=Q\cup R\cup S\cup\dots$.

Before leaving this short analysis for the gravitational part, we have to remark that the terms that contain the $E_P$ factors are purely contact terms, now absorbed in the field redefinition.

The field on the loop will be the Proca field denoted by $A_{\mu}$. The full action of gravity coupled to a massive vector field is the following
\begin{eqnarray}
S_{\text{GR+A}}=S_{\text{LL}}+S_{\text{P}},
\end{eqnarray}
where $S_{\text{P}}$ is the part that contains the interaction with $A_{\mu}$, explicitly
\begin{eqnarray}\label{eq:action-proca}
S_{\mathrm{P}}&=&\int d^{4}x\,\big(-\tfrac{1}{4}(-\mathfrak{g})^{-1/2}\mathfrak{g}^{\mu\nu}\mathfrak{g}^{\rho\sigma}F_{\mu\rho}F_{\nu\sigma}+\tfrac{m^{2}}{2}\mathfrak{g}^{\mu\nu}A_{\mu}A_{\nu}\big)\notag\\
&=&\int d^{4}x\,\big(-\tfrac{1}{2}(-\mathfrak{g})^{-1/2}\mathfrak{g}^{\mu\nu}\mathfrak{g}^{\rho\sigma}(\partial_{\mu}A_{\rho}\partial_{\nu}A_{\sigma}-\partial_{\rho}A_{\mu}\partial_{\nu}A_{\sigma})+\tfrac{m^{2}}{2}\mathfrak{g}^{\mu\nu}A_{\mu}A_{\nu}\big).
\end{eqnarray}
The first simplification we have on this approach is in the mass term, while the \emph{Maxwell term} retains the same form as when coupled to the metric $g^{\mu\nu}$.

The equations of motion for the Proca field read
\begin{equation}\label{eq:eom-proca}
\mathfrak{g}^{\mu\nu}\partial_{\nu}[(-\mathfrak{g})^{-1/2}\mathfrak{g}^{\rho\alpha}(\partial_{\mu}A_{\rho}-\partial_{\rho}A_{\mu})]+m^{2}\mathfrak{g}^{\mu\alpha}A_{\mu}=0
\end{equation}

Together with \eqref{eq:forsol-g} we replace the formal solution for the vector field 
\begin{equation}\label{eq:forsol-a}
A_{\mu}(x)=\sum_{P}\mathcal{A}_{P\mu}e^{ik_{P}\cdot x},
\end{equation}
with $\mathcal{A}_{k\mu}=\varepsilon_{k\mu}$ the vector field polarization. Additionally, we have an expansion for the $(-\mathfrak{g})^{-1/2}$ in terms of multi-particle fields, that we write as
\begin{eqnarray}\label{eq:detg}
(-\mathfrak{g})^{-1/2}=1+\sum_{P}E_Pe^{ik_{P}\cdot x}
\end{eqnarray}
where the $E_P$ are the terms that come from the known expansion of the square root of the determinant
\begin{eqnarray}\label{eq:detg-pert}
E_1 &=& 0,\\
E_{12} &=&-\tfrac{1}{2}\eta_{\mu\nu}I_{12}^{\mu\nu}+\tfrac{1}{2}\eta_{\mu\rho}\eta_{\nu\sigma}h_1^{\mu\nu}h_{2}^{\rho\sigma},\\
&& \vdots\notag
\end{eqnarray}

After replacing \eqref{eq:forsol-g}, \eqref{eq:forsol-a}, and \eqref{eq:detg} on \eqref{eq:eom-proca}; we solve and find the recursive expression for the Berends-Giele currents for the Proca field
\begin{eqnarray}\label{eq:pert-a}
(s_{P}-m^{2})\mathcal{A}_{P\beta}&=&\sum_{P=Q\cup R}\Big\{[E_{R}\eta^{\rho\alpha}\eta^{\mu\nu}-\eta^{\rho\alpha}I_{R}^{\mu\nu}-I_{R}^{\rho\alpha}\eta^{\mu\nu}][V_{P,Q}]_{\beta\alpha\mu\nu\rho}^{\sigma}\mathcal{A}_{Q\sigma}\notag\\
&&+(k_{P\beta}k_{Q\alpha}-m^2\eta_{\beta\alpha})I_{R}^{\alpha\sigma}\mathcal{A}_{Q\sigma}\Big\}\notag\\
&&+\sum_{P=Q\cup R\cup S}\Big\{[I_{S}^{\rho\alpha}I_{R}^{\mu\nu}-E_{S}\eta^{\rho\alpha}I_{R}^{\mu\nu}-E_{S}I_{R}^{\rho\alpha}\eta^{\mu\nu}][V_{P,Q}]_{\beta\alpha\mu\nu\rho}^{\sigma}\mathcal{A}_{Q\sigma}\Big\}\notag\\
&&+\sum_{P=Q\cup R\cup S\cup T}\Big\{ E_{T}I_{S}^{\rho\alpha}I_{R}^{\mu\nu}[V_{P,Q}]_{\beta\alpha\mu\nu\rho}^{\sigma}\mathcal{A}_{Q\sigma}\Big\}
\end{eqnarray}
where
\begin{eqnarray}
[V_{P,Q}]_{\alpha\beta\mu\nu\rho}^{\sigma}&=&(\tfrac{k_{P\alpha}k_{P\beta}}{m^{2}}-\eta_{\beta\alpha})k_{P\nu}(\delta_{\rho}^{\sigma}k_{Q\mu}-\delta_{\mu}^{\sigma}k_{Q\rho}).
\end{eqnarray}

In order to have and off-shell Proca field, the relation \eqref{eq:pert-a} does not hold when $P$ is a single index word $P=p$. We see that \eqref{eq:pert-a} has an additional off-shell leg, the propagator on the left hand side, that we can use to close the loop. To obtain the integrands we will now follow a procedure close to the one from \cite{Gomez:2022dzk} that we will explain now.

Aiming at the one-loop integrand, we start by factorizing the loop-closing leg $\mathcal{A}_{\ell P\beta}=\varepsilon_{\ell\alpha}\mathcal{J}_{P\beta}^{\sigma}(\ell)$ in \eqref{eq:pert-a} and take the word to decompose as $\tilde{P}=\ell P$, here $\varepsilon_{\ell\alpha}$ being the dummy polarization to effectively close the loop. Our approach here has a slightly variation from the one \cite{Gomez:2022dzk}, in the sense that after making the aforementioned replacement in the right hand side of  \eqref{eq:pert-a}, we keep only the terms without $\mathcal{J}_{P\beta}^{\sigma}(\ell)$ on them. After all this we arrive at our one-loop integrands building blocks that read 
\begin{eqnarray}\label{eq:loop-curr}
((\ell+k_{P})^{2}-m^{2})\mathcal{J}_{P\beta}^{\sigma}(\ell)&=&[E_{P}\eta^{\rho\alpha}\eta^{\mu\nu}-\eta^{\rho\alpha}I_{P}^{\mu\nu}-I_{P}^{\rho\alpha}\eta^{\mu\nu}][V_{\ell P,\ell}]_{\beta\alpha\mu\nu\rho}^{\sigma}\notag\\
&&+((\ell+k_{P})_{\beta}\ell_{\alpha}-m^{2}\eta_{\beta\alpha})I_{P}^{\alpha\sigma}\notag\\
&&+\sum_{P=Q\cup R}[I_{Q}^{\rho\alpha}I_{R}^{\mu\nu}-E_{Q}\eta^{\rho\alpha}I_{R}^{\mu\nu}-E_{Q}I_{R}^{\rho\alpha}\eta^{\mu\nu}][V_{\ell P,\ell}]_{\beta\alpha\mu\nu\rho}^{\sigma}\notag\\
&&+\sum_{P=Q\cup R\cup S}E_{Q}I_{S}^{\rho\alpha}I_{R}^{\mu\nu}[V_{\ell P,\ell}]_{\beta\alpha\mu\nu\rho}^{\sigma}.
\end{eqnarray}

These objects have a distribution of $|P|$ external particles attached at one point and two off-shell legs to be sewn and form a loop. We can represent them pictorially for a better understanding
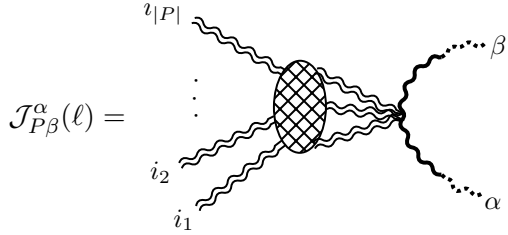
\begin{equation*}
\mathcal{J}_{P\beta}^{\alpha}(\ell)=
\tikzset{
pattern size/.store in=\mcSize, 
pattern size = 5pt,
pattern thickness/.store in=\mcThickness, 
pattern thickness = 0.3pt,
pattern radius/.store in=\mcRadius, 
pattern radius = 1pt}
\makeatletter
\pgfutil@ifundefined{pgf@pattern@name@_4rncbadvf}{
\pgfdeclarepatternformonly[\mcThickness,\mcSize]{_4rncbadvf}
{\pgfqpoint{0pt}{0pt}}
{\pgfpoint{\mcSize}{\mcSize}}
{\pgfpoint{\mcSize}{\mcSize}}
{
\pgfsetcolor{\tikz@pattern@color}
\pgfsetlinewidth{\mcThickness}
\pgfpathmoveto{\pgfqpoint{0pt}{\mcSize}}
\pgfpathlineto{\pgfpoint{\mcSize+\mcThickness}{-\mcThickness}}
\pgfpathmoveto{\pgfqpoint{0pt}{0pt}}
\pgfpathlineto{\pgfpoint{\mcSize+\mcThickness}{\mcSize+\mcThickness}}
\pgfusepath{stroke}
}}
\makeatother
\tikzset{every picture/.style={line width=0.75pt}} 
\begin{tikzpicture}[x=0.75pt,y=0.75pt,yscale=-1,xscale=1,baseline=-5.7cm]

\draw    (474.33,257.05) .. controls (474.64,254.71) and (475.96,253.7) .. (478.3,254.01) .. controls (480.63,254.32) and (481.95,253.3) .. (482.26,250.97) .. controls (482.57,248.63) and (483.89,247.61) .. (486.23,247.92) .. controls (488.56,248.23) and (489.89,247.21) .. (490.2,244.88) .. controls (490.51,242.55) and (491.84,241.53) .. (494.17,241.84) .. controls (496.5,242.15) and (497.83,241.13) .. (498.14,238.8) .. controls (498.44,236.46) and (499.76,235.44) .. (502.1,235.75) .. controls (504.43,236.06) and (505.76,235.04) .. (506.07,232.71) .. controls (506.38,230.38) and (507.71,229.36) .. (510.04,229.67) .. controls (512.37,229.98) and (513.7,228.96) .. (514.01,226.63) -- (517.67,223.82) -- (517.67,223.82)(476.15,259.43) .. controls (476.46,257.1) and (477.79,256.08) .. (480.12,256.39) .. controls (482.45,256.7) and (483.78,255.68) .. (484.09,253.35) .. controls (484.4,251.01) and (485.72,249.99) .. (488.06,250.3) .. controls (490.39,250.61) and (491.71,249.59) .. (492.02,247.26) .. controls (492.33,244.93) and (493.66,243.91) .. (495.99,244.22) .. controls (498.32,244.53) and (499.65,243.51) .. (499.96,241.18) .. controls (500.27,238.84) and (501.59,237.82) .. (503.93,238.13) .. controls (506.26,238.44) and (507.59,237.42) .. (507.9,235.09) .. controls (508.21,232.76) and (509.53,231.74) .. (511.86,232.05) .. controls (514.19,232.36) and (515.52,231.34) .. (515.83,229.01) -- (519.49,226.2) -- (519.49,226.2) ;
\draw [line width=1.5]    (598.04,241.12) .. controls (595.43,241.45) and (593.97,240.48) .. (593.65,238.21) .. controls (593.26,235.8) and (592,234.77) .. (589.87,235.13) .. controls (587.44,235.12) and (586.27,233.92) .. (586.34,231.54) .. controls (586.65,229.29) and (585.62,227.92) .. (583.23,227.43) .. controls (581,226.98) and (580.22,225.57) .. (580.88,223.21) .. controls (581.79,221.07) and (581.21,219.51) .. (579.13,218.53) .. controls (577.11,217.3) and (576.77,215.61) .. (578.11,213.46) .. controls (579.65,211.84) and (579.58,210.18) .. (577.91,208.47) .. controls (576.36,206.79) and (576.55,205.19) .. (578.46,203.66) .. controls (580.45,202.43) and (580.89,200.8) .. (579.78,198.77) .. controls (578.86,196.56) and (579.52,195.07) .. (581.76,194.29) .. controls (583.96,193.82) and (584.87,192.41) .. (584.48,190.07) .. controls (584.38,187.62) and (585.53,186.4) .. (587.94,186.39) .. controls (590.15,186.77) and (591.53,185.83) .. (592.06,183.57) .. controls (593.08,181.35) and (594.63,180.84) .. (596.72,182.05) -- (598.04,181.92) ;
\draw [line width=1.5]  [dash pattern={on 1.5pt off 1.5pt}]  (618.04,250.88) .. controls (616.02,252.19) and (614.37,251.88) .. (613.09,249.95) .. controls (611.79,248.01) and (610.14,247.68) .. (608.13,248.97) .. controls (605.9,250.02) and (604.44,249.31) .. (603.75,246.86) .. controls (603.8,244.51) and (602.64,243.3) .. (600.26,243.24) -- (598.04,241.12) ;
\draw [line width=1.5]  [dash pattern={on 1.5pt off 1.5pt}]  (619.64,175.12) .. controls (618.03,176.96) and (616.33,177.11) .. (614.54,175.58) .. controls (612.83,173.99) and (611.17,174.01) .. (609.57,175.64) .. controls (607.68,177.06) and (606.19,177.38) .. (605.12,176.6) .. controls (602.74,176.39) and (601.41,177.43) .. (601.14,179.72) -- (598.04,181.92) ;
\draw [line width=0.75]    (535.27,187.68) .. controls (537.52,186.97) and (539,187.74) .. (539.71,189.99) .. controls (540.42,192.24) and (541.89,193.01) .. (544.14,192.3) .. controls (546.39,191.59) and (547.87,192.36) .. (548.58,194.61) .. controls (549.29,196.86) and (550.76,197.62) .. (553.01,196.91) .. controls (555.26,196.2) and (556.74,196.97) .. (557.45,199.22) .. controls (558.16,201.47) and (559.63,202.24) .. (561.88,201.53) .. controls (564.13,200.82) and (565.61,201.59) .. (566.32,203.84) .. controls (567.03,206.09) and (568.51,206.85) .. (570.76,206.14) .. controls (573.01,205.43) and (574.48,206.2) .. (575.19,208.45) -- (579.45,210.67) -- (579.45,210.67)(533.89,190.34) .. controls (536.14,189.63) and (537.61,190.4) .. (538.32,192.65) .. controls (539.03,194.9) and (540.51,195.67) .. (542.76,194.96) .. controls (545.01,194.25) and (546.48,195.02) .. (547.19,197.27) .. controls (547.9,199.52) and (549.38,200.28) .. (551.63,199.57) .. controls (553.88,198.86) and (555.35,199.63) .. (556.06,201.88) .. controls (556.77,204.13) and (558.25,204.9) .. (560.5,204.19) .. controls (562.75,203.48) and (564.23,204.25) .. (564.94,206.5) .. controls (565.65,208.75) and (567.12,209.52) .. (569.37,208.81) .. controls (571.62,208.1) and (573.1,208.86) .. (573.81,211.11) -- (578.07,213.33) -- (578.07,213.33) ;
\draw  [pattern=_4rncbadvf,pattern size=6pt,pattern thickness=0.75pt,pattern radius=0pt, pattern color={rgb, 255:red, 0; green, 0; blue, 0}] (526.79,183.7) .. controls (534.04,183.75) and (539.85,194.18) .. (539.78,207.01) .. controls (539.7,219.84) and (533.77,230.21) .. (526.52,230.17) .. controls (519.27,230.13) and (513.46,219.69) .. (513.53,206.86) .. controls (513.61,194.03) and (519.54,183.66) .. (526.79,183.7) -- cycle ;
\draw    (533.32,225.19) .. controls (534.39,223.09) and (535.97,222.57) .. (538.07,223.64) .. controls (540.17,224.71) and (541.76,224.2) .. (542.83,222.1) .. controls (543.9,220) and (545.48,219.48) .. (547.58,220.55) .. controls (549.68,221.62) and (551.27,221.11) .. (552.34,219.01) .. controls (553.41,216.91) and (554.99,216.39) .. (557.09,217.46) .. controls (559.19,218.53) and (560.78,218.02) .. (561.85,215.92) .. controls (562.92,213.82) and (564.5,213.3) .. (566.6,214.37) .. controls (568.7,215.44) and (570.29,214.93) .. (571.36,212.83) .. controls (572.43,210.73) and (574.01,210.21) .. (576.11,211.28) -- (578.3,210.57) -- (578.3,210.57)(534.24,228.04) .. controls (535.31,225.94) and (536.9,225.42) .. (539,226.49) .. controls (541.1,227.56) and (542.68,227.05) .. (543.75,224.95) .. controls (544.82,222.85) and (546.41,222.33) .. (548.51,223.4) .. controls (550.61,224.47) and (552.19,223.96) .. (553.26,221.86) .. controls (554.33,219.76) and (555.92,219.25) .. (558.02,220.32) .. controls (560.12,221.39) and (561.7,220.87) .. (562.77,218.77) .. controls (563.84,216.67) and (565.43,216.16) .. (567.53,217.23) .. controls (569.63,218.3) and (571.21,217.78) .. (572.28,215.68) .. controls (573.35,213.58) and (574.94,213.07) .. (577.04,214.14) -- (579.22,213.43) -- (579.22,213.43) ;
\draw [line width=0.75]    (539.97,205.52) .. controls (541.84,204.08) and (543.49,204.29) .. (544.93,206.16) .. controls (546.37,208.03) and (548.02,208.24) .. (549.89,206.79) .. controls (551.76,205.35) and (553.41,205.56) .. (554.85,207.43) .. controls (556.29,209.3) and (557.94,209.51) .. (559.81,208.06) .. controls (561.68,206.62) and (563.33,206.83) .. (564.77,208.7) .. controls (566.21,210.57) and (567.86,210.78) .. (569.73,209.33) .. controls (571.6,207.89) and (573.25,208.1) .. (574.69,209.97) -- (578.95,210.51) -- (578.95,210.51)(539.59,208.5) .. controls (541.46,207.06) and (543.11,207.27) .. (544.55,209.14) .. controls (545.99,211.01) and (547.64,211.22) .. (549.51,209.77) .. controls (551.38,208.32) and (553.03,208.53) .. (554.47,210.4) .. controls (555.91,212.27) and (557.56,212.48) .. (559.43,211.04) .. controls (561.3,209.59) and (562.95,209.8) .. (564.39,211.67) .. controls (565.83,213.54) and (567.48,213.75) .. (569.35,212.31) .. controls (571.22,210.86) and (572.87,211.07) .. (574.31,212.94) -- (578.57,213.49) -- (578.57,213.49) ;
\draw    (465.77,235.3) .. controls (466.53,233.07) and (468.03,232.33) .. (470.26,233.08) .. controls (472.49,233.83) and (473.99,233.09) .. (474.74,230.86) .. controls (475.49,228.63) and (476.99,227.89) .. (479.22,228.64) .. controls (481.45,229.39) and (482.95,228.65) .. (483.7,226.42) .. controls (484.45,224.19) and (485.95,223.45) .. (488.18,224.21) .. controls (490.41,224.96) and (491.91,224.22) .. (492.66,221.99) .. controls (493.41,219.76) and (494.91,219.02) .. (497.14,219.77) .. controls (499.37,220.52) and (500.87,219.78) .. (501.62,217.55) .. controls (502.38,215.32) and (503.88,214.58) .. (506.11,215.33) .. controls (508.34,216.09) and (509.84,215.35) .. (510.59,213.12) -- (513.51,211.67) -- (513.51,211.67)(467.11,237.98) .. controls (467.86,235.75) and (469.36,235.02) .. (471.59,235.77) .. controls (473.82,236.52) and (475.32,235.78) .. (476.07,233.55) .. controls (476.82,231.32) and (478.32,230.58) .. (480.55,231.33) .. controls (482.78,232.08) and (484.28,231.34) .. (485.03,229.11) .. controls (485.78,226.88) and (487.28,226.14) .. (489.51,226.89) .. controls (491.74,227.65) and (493.24,226.91) .. (493.99,224.68) .. controls (494.74,222.45) and (496.24,221.71) .. (498.47,222.46) .. controls (500.7,223.21) and (502.2,222.47) .. (502.95,220.24) .. controls (503.71,218.01) and (505.21,217.27) .. (507.44,218.02) .. controls (509.67,218.78) and (511.17,218.04) .. (511.92,215.81) -- (514.84,214.36) -- (514.84,214.36) ;
\draw    (473.59,162.14) .. controls (475.86,161.52) and (477.3,162.35) .. (477.92,164.62) .. controls (478.54,166.89) and (479.99,167.72) .. (482.26,167.11) .. controls (484.53,166.49) and (485.98,167.32) .. (486.6,169.59) .. controls (487.22,171.86) and (488.67,172.69) .. (490.94,172.08) .. controls (493.21,171.47) and (494.66,172.3) .. (495.28,174.57) .. controls (495.9,176.84) and (497.35,177.67) .. (499.62,177.05) .. controls (501.89,176.44) and (503.34,177.27) .. (503.96,179.54) .. controls (504.57,181.81) and (506.02,182.64) .. (508.29,182.02) .. controls (510.56,181.41) and (512.01,182.24) .. (512.63,184.51) .. controls (513.25,186.78) and (514.7,187.61) .. (516.97,186.99) -- (518.92,188.11) -- (518.92,188.11)(472.09,164.74) .. controls (474.37,164.13) and (475.82,164.96) .. (476.43,167.23) .. controls (477.05,169.5) and (478.5,170.33) .. (480.77,169.71) .. controls (483.04,169.1) and (484.49,169.93) .. (485.11,172.2) .. controls (485.73,174.47) and (487.18,175.3) .. (489.45,174.68) .. controls (491.72,174.07) and (493.17,174.9) .. (493.79,177.17) .. controls (494.41,179.44) and (495.86,180.27) .. (498.13,179.65) .. controls (500.4,179.04) and (501.85,179.87) .. (502.46,182.14) .. controls (503.08,184.41) and (504.53,185.24) .. (506.8,184.62) .. controls (509.07,184.01) and (510.52,184.84) .. (511.14,187.11) .. controls (511.76,189.38) and (513.21,190.21) .. (515.48,189.6) -- (517.43,190.71) -- (517.43,190.71) ;

\draw (626.72,255.96) node  [font=\small] [align=left] {\begin{minipage}[lt]{10.5pt}\setlength\topsep{0pt}
$\displaystyle \alpha $
\end{minipage}};
\draw (629.52,176.76) node  [font=\small] [align=left] {\begin{minipage}[lt]{10.5pt}\setlength\topsep{0pt}
$\displaystyle \beta $
\end{minipage}};
\draw (469.92,263.96) node  [font=\small] [align=left] {\begin{minipage}[lt]{10.5pt}\setlength\topsep{0pt}
$\displaystyle i_{1}$
\end{minipage}};
\draw (458.12,239.76) node  [font=\small] [align=left] {\begin{minipage}[lt]{10.5pt}\setlength\topsep{0pt}
$\displaystyle i_{2}$
\end{minipage}};
\draw (454.72,158.76) node  [font=\small] [align=left] {\begin{minipage}[lt]{10.5pt}\setlength\topsep{0pt}
$\displaystyle i_{|P|}$
\end{minipage}};
\draw (478.52,211.96) node  [font=\small] [align=left] {\begin{minipage}[lt]{9.14pt}\setlength\topsep{0pt}
$\displaystyle \cdot $
\end{minipage}};
\draw (477.72,203.16) node  [font=\small] [align=left] {\begin{minipage}[lt]{9.14pt}\setlength\topsep{0pt}
$\displaystyle \cdot $
\end{minipage}};
\draw (478.52,193.96) node  [font=\small] [align=left] {\begin{minipage}[lt]{9.14pt}\setlength\topsep{0pt}
$\displaystyle \cdot $
\end{minipage}};
\end{tikzpicture}
\end{equation*}

The $\mathcal{J}_{P\beta}^{\sigma}(\ell)$ give us the integrand for a tadpole if we take its trace, and a cyclic contractions of $n$ of them will give us the integrand for an $n$-gon with external trees (each tree with $P_i$ particle content) attached, namely
\begin{eqnarray}\label{eq:treen-gon}
I^J(P_1,P_2,P_3,\dots,P_n;\ell)=\mathcal{J}_{P_1\beta_1}^{\alpha}(\ell)\mathcal{J}_{P_2\beta_2}^{\beta_1}(\ell+k_{P_1})\mathcal{J}_{P_3\beta_3}^{\beta_2}(\ell+k_{P_1P_2})\dots\mathcal{J}_{P_N\alpha}^{\beta_{n-1}}(\ell-k_{P_n}),\notag\\
\end{eqnarray} 
where we have imposed momentum conservation $k^{\mu}_{P_1P_2\dots P_n}=0$. The expression \eqref{eq:treen-gon} also includes the contact terms, as we will see later.

A noteworthy aspect of the method used to obtain the integrands in this work, beyond relying solely on the classical equations of motion, is its efficiency in handling the combinatorial complexity typically associated with Feynman diagrams. In the present four-point case, for instance, one would normally encounter six distinct diagram topologies and their permutations to compute the full amplitude. In contrast, our approach requires only the calculation of $\mathcal{J}_{P}(\ell)$, whose form depends on the specific theory under consideration. For any one-loop four-point amplitude, all contributions can then be obtained by summing over the following three basic structures and their respective permutations.
\begin{eqnarray}
I^J(1,2,3,4;\ell)&=&\mathcal{J}_{1}(\ell)\cdot\mathcal{J}_{2}(\ell+k_1)\cdot\mathcal{J}_{3}(\ell+k_{12})\cdot\mathcal{J}_{4}(\ell-k_4),\\
I^J(12,3,4;\ell)&=&\mathcal{J}_{12}(\ell)\cdot\mathcal{J}_{3}(\ell+k_{12})\cdot\mathcal{J}_{4}(\ell-k_4),\\
I^J(12,34;\ell)&=&\mathcal{J}_{12}(\ell)\cdot\mathcal{J}_{34}(\ell+k_{12}),
\end{eqnarray} 
where the ``$\,\cdot\,$'' refers to the type of contraction depending on the nature of $\mathcal{J}_{P}(\ell)$ and the number of permutations to be taken depend if the amplitude is ordered or not. Additionally, we can simplify the building blocks at several steps of the process to obtain the simplest integrands and remove unwanted singularities.  

Back to our case of study, from \eqref{eq:treen-gon} we obtain the four-point integrands by restricting $|P_1P_2\dots P_n|=4$ and $|P_i|\leq 2$, the latter condition takes \eqref{eq:loop-curr} to the form
\begin{eqnarray}\label{eq:intg-gr}
((\ell+k_{P})^{2}-m^{2})\mathcal{J}_{P\beta}^{\sigma}(\ell)&=&I_{P}^{\mu\nu}\ell_{\mu}\ell_{\nu}\delta_{\beta}^{\sigma}+k_{P\beta}I_{P}^{\alpha\sigma}\ell_{\alpha}-\eta_{\beta\alpha}I_{P}^{\rho\alpha}\ell_{\rho}(\ell+k_{P})^{\sigma}\notag\\
&&+E_{P}(\ell_{\beta}(\ell+k_{P})^{\sigma}-\delta_{\beta}^{\sigma}(\ell+k_{P})\cdot\ell)\notag\\
&&+\tfrac{1}{2}(((\ell+k_{P})^{2}-m^{2})+(\ell^{2}-m^{2})-s_{P})\eta_{\beta\alpha}I_{P}^{\alpha\sigma}\notag\\
&&-\sum_{P=Q\cup R}\eta_{\beta\alpha}(\ell+k_{Q})_{\nu}\ell_{\rho}(I_{Q}^{\sigma\alpha}I_{R}^{\rho\nu}-I_{Q}^{\rho\alpha}I_{R}^{\sigma\nu}).
\end{eqnarray}
Notice that here the mass appears only into inverse propagators, this is a simplification brought by the metric density approach.

The expression \eqref{eq:intg-gr} gives all the simplifications for the integrands, but we will not use it to calculate our amplitudes. Before going to the integration, we want to see what would be the effect of assuming a double copy factorization for the external trees. In order to analyse the structure of the trees attached to the loop let us briefly review some aspects of tree-level double copy in the next section. 

\section{Tree-level double copy for the currents}

Let us now analyse the structure of the external gravitational trees, here from a double copy perspective to see later its effect on the one-loop amplitudes. In \cite{Ahmadiniaz:2021ayd} it was shown that the gravitational currents admit a double copy representation, without the need of a Kawai-Lewellen-Tye kernel like is presented for other theories in \cite{Mizera:2018jbh}. Borrowing the notation from \cite{Ahmadiniaz:2021ayd}, we write the gravitational currents in the double copy prescription as follows  
\begin{equation}\label{eq:current-dc}
\mathcal{G}^{\mu\nu}_P = \llbracket \bar{\varepsilon}^{\mu} \otimes \varepsilon^{\nu} \rrbracket \circ b_{\cd}(P). 
\end{equation}
Here $b_{\cd}$ is a map inspired on the \emph{binary tree map}, originally introduced in \cite{Bridges:2019siz}, but was adapted to colour-dressed currents (therefore the subscript $\cd$). Such a map acts on ordered words and it is recursively defined by
\begin{align}\label{eq:cdBGmap}
\begin{split}
b_{\cd}(i) &= i, \\
b_{\cd}(P) &= \frac{1}{2 s_P} \sum_{P =  Q \cup R} [b_{\cd}(Q),b_{\cd}(R)].
\end{split}
\end{align}
Notice that we also changed the name for the gravitational currents on \eqref{eq:current-dc} to differentiate them from the currents $I_P^{\mu\nu}$ for the metric density.
   
To see how \eqref{eq:current-dc} works we present some examples
\begin{align}\label{eq:DC-BG23}
\begin{split}
\mathcal{G}^{\mu\nu}_{1} &= \bar{\varepsilon}^{\mu}_1 \varepsilon^{\nu}_{1}, \\
\mathcal{G}_{12}^{\mu\nu} &= \frac{\bar{\varepsilon}^{\mu}_{[1,2]} \varepsilon^{\nu}_{[1,2]}}{s_{12}}, \\
\mathcal{G}_{123}^{\mu\nu} &= \frac{\bar{\varepsilon}^{\mu}_{[[1,2],3]} \varepsilon^{\nu}_{[[1,2],3]}}{s_{12} s_{123}} +  \frac{\bar{\varepsilon}^{\mu}_{[[1,3],2]} \varepsilon^{\nu}_{[[1,3],2]}}{s_{13} s_{123}} +  \frac{\bar{\varepsilon}^{\mu}_{[[2,3],1]} \varepsilon^{\nu}_{[[2,3],1]}}{s_{23} s_{123}}.
\end{split}
\end{align}
The $\varepsilon_{[P]}^{\mu}$ are the numerators of the pure Yang-Mills currents in the colour-kinematics gauge \cite{Mafra:2014oia}, meaning that they satisfy the same Jacobi identities satisfied by their respective colour factor in the Yang-Mills amplitude. The structure of the colour factor consist of nested commutators of the gauge group generators, and this is explicitly displayed on their multi-particle indices bracket structure, i.e.  
\begin{align}
\begin{split}
&\varepsilon^{\mu}_{[1,2]} + \varepsilon^{\mu}_{[2,1]}=0,\\
&\varepsilon^{\mu}_{[[1,2],3]} + \varepsilon^{\mu}_{[[1,2],3]}=0,\quad\varepsilon^{\mu}_{[[1,2],3]}+\varepsilon^{\mu}_{[[3,1],2]}+\varepsilon^{\mu}_{[[2,3],1]}=0.
\end{split}
\end{align}

For our four-point calculation we will only need the rank-2 current whose numerator we write explicitly here
\begin{eqnarray}
\varepsilon^{\mu}_{[1,2]}=(\varepsilon_1\cdot k_2)\varepsilon_2^{\mu}-(\varepsilon_2\cdot k_1)\varepsilon_1^{\mu}+ \tfrac{1}{2}(\varepsilon_1\cdot\varepsilon_2)(k_1-k_2)^{\mu}.
\end{eqnarray}

A general expression for any rank was found in \cite{Bridges:2019siz}, later in \cite{Ahmadiniaz:2021fey} it was shown that this gauge arise naturally from the pinching procedure in the Bern-Kosower formalism for gluon amplitudes \cite{Bern1991-1669}. In the practice we take for the single particle polarizations $\bar{\varepsilon}^{\mu}_i=\varepsilon^{\mu}_i$ and $\varepsilon_i\cdot\varepsilon_i=0$, this in order to be consistent with the graviton being traceless. 

Even at tree-level we can start to see some of the advantages double copy brings. Apart from the obvious fact that we do not need gravity at all, many expressions become simpler. for example, if we use the current that comes directly from gravity in $D$-dimensions, for \eqref{eq:detg-pert} we have the explicit expression
\begin{eqnarray}
E_{12}=\frac{10-D}{8}(\varepsilon_1\cdot\varepsilon_2)^2 + \frac{D-2}{4}\frac{(\varepsilon_1\cdot\varepsilon_2)(\varepsilon_1\cdot k_2)(\varepsilon_2\cdot k_1)}{k_1\cdot k_2},
\end{eqnarray}
while from double copy, also in $D$-dimensions, we get
\begin{eqnarray}
\tilde{E}_{12}=\frac{5}{8}(\varepsilon_1\cdot\varepsilon_2)^2.
\end{eqnarray}

Both expressions are the same at $D=4$, after suitable choice of reference momenta for the polarizations (i.e. both are zero for the same helicities or lead to the same expression for a product of polarizations with different helicities), but we save additional steps starting from the double copy version. There will be another advantages that we will mention once we have the expressions for the one-loop integrands, which are the subject of the next section.

\section{Implications of tree-level double copy on the one-loop amplitude}

We go back to the four-graviton amplitude with a Proca field on the loop. The contributions for the four-graviton amplitude are three boxes, six triangles and two bubbles, we will analyse each type of contribution. Starting with the box integrand, from \eqref{eq:loop-curr} and \eqref{eq:treen-gon} we have
\begin{eqnarray}
((\ell+k_{1})^{2}-m^{2})\mathcal{J}_{1\beta}^{\sigma}(\ell)&=&h_{1}^{\mu\nu}\ell_{\mu}\ell_{\nu}\delta_{\beta}^{\sigma}+k_{1\beta}h_{1}^{\alpha\sigma}\ell_{\alpha}-\eta_{\beta\alpha}h_{1}^{\rho\alpha}\ell_{\rho}(\ell+k_{1})^{\sigma}\notag\\&&+\tfrac{1}{2}(((\ell+k_{1})^{2}-m^{2})+(\ell^{2}-m^{2}))\eta_{\beta\alpha}h_{1}^{\alpha\sigma}.
\end{eqnarray}
The first line has only cubic interactions. In the second line we have a sum of two inverse propagators that will cancel the respective factor in the numerator and pinch adjacent external legs, thus giving us contact terms that will contribute to the triangle and bubble configurations with quartic interactions. In the next step we write the single graviton particle as $h_1^{\mu\nu}=\varepsilon_1^{\mu}\varepsilon_1^{\nu}$, and this replacement allow us to see that the factors that will compose the box numerator also admit a factorization as we see now
\begin{eqnarray}
((\ell+k_{1})^{2}-m^{2})J_{1\beta}^{\alpha}(\ell)=(\varepsilon_{1}\cdot\ell)(\varepsilon_{1}\cdot\ell\delta_{\beta}^{\alpha}-\varepsilon_{1\beta}(\ell+k_{1})^{\alpha}+k_{1\beta}\varepsilon_{1}^{\alpha}),
\end{eqnarray}
with the first factor being completely scalar and the one with a vector structure contained only on the second factor. This is more evident to see after we write the whole numerator for the box, which has the following structure
\begin{eqnarray}
n^{\text{GR+A}}(1,2,3,4;\ell) = n^{[0]}(1,2,3,4;\ell)n^{[1]}(1,2,3,4;\ell)
\end{eqnarray}
with
\begin{eqnarray}
n^{[0]}(1,2,3,4;\ell) &=& (\varepsilon_{1}\cdot\ell)(\varepsilon_{2}\cdot(\ell+k_{1}))(\varepsilon_{3}\cdot(\ell+k_{12}))(\varepsilon_{4}\cdot\ell),\\
n^{[1]}(1,2,3,4;\ell) &=& (\varepsilon_{1}\cdot\ell\delta_{\beta}^{\alpha}-\varepsilon_{1\beta}(\ell+k_{1})^{\alpha}+k_{1\beta}\varepsilon_{1}^{\alpha})(\varepsilon_{2}\cdot(\ell+k_{1})\delta_{\gamma}^{\beta}-\varepsilon_{2\gamma}(\ell+k_{12})^{\beta}+k_{2\gamma}\varepsilon_{2}^{\beta})\notag\\
&&\times(\varepsilon_{3}\cdot(\ell+k_{12})\delta_{\lambda}^{\gamma}-\varepsilon_{3\lambda}(\ell+k_{123})^{\gamma}+k_{3\lambda}\varepsilon_{3}^{\gamma})(\varepsilon_{4}\cdot\ell\delta_{\alpha}^{\lambda}-\varepsilon_{4\alpha}\ell^{\lambda}+k_{4\alpha}\varepsilon_{4}^{\lambda}).\notag\\
\end{eqnarray}

Clearly we have a double copy-like factorization here, but the numerators $n^{[0]}(1,2,3,4;\ell)$ and $n^{[1]}(1,2,3,4;\ell)$ are not the so-called one-loop BCJ master numerators \cite{Bern:2013yya,Nohle:2013bfa}. This means that they do not satisfy colour-kinematics duality, a property that cannot be expected, since the particle in the loop carries no color structure. In this setup, the ``master'' numerators are instead determined directly from the simpler abelian interactions with the metric density, rather than being constructed to obey Jacobi-like relations. However, the same type of factorization naturally emerges once we take into account the tree-level double-copy structure for the external trees from \eqref{eq:DC-BG23}. 

For the case of the triangle contributions we need the building block of rank-2, that we also simplify in order to avoid tadpole integrals. It takes the following form
\begin{eqnarray}
((\ell+k_{12})^{2}-m^{2})\mathcal{J}_{12\beta}^{\alpha}(\ell)&=&I_{12}^{\mu\nu}\ell_{\mu}\ell_{\nu}\delta_{\beta}^{\sigma}+k_{12\beta}I_{12}^{\alpha\sigma}\ell_{\alpha}-\eta_{\beta\alpha}I_{12}^{\rho\alpha}\ell_{\rho}(\ell+k_{P})^{\sigma}\notag\\
&&+\tfrac{1}{2}s_{12}(E_{12}\delta_{\beta}^{\alpha}-\eta_{\beta\sigma}I_{12}^{\sigma\alpha})\notag\\
&&-(\ell+k_{1})\cdot\varepsilon_{2}\varepsilon_{1\beta}(\varepsilon_{1}^{\alpha}\ell\cdot\varepsilon_{2}-\varepsilon_{2}^{\alpha}\ell\cdot\varepsilon_{1})\notag\\
&&-(\ell+k_{2})\cdot\varepsilon_{1}\varepsilon_{2\beta}(\varepsilon_{2}^{\alpha}\ell\cdot\varepsilon_{1}-\varepsilon_{1}^{\alpha}\ell\cdot\varepsilon_{2}).
\end{eqnarray}
We now isolate the part with cubic interactions only
\begin{eqnarray}
((\ell+k_{12})^{2}-m^{2})J_{12\beta}^{\sigma}(\ell)=I_{12}^{\mu\nu}\ell_{\mu}\ell_{\nu}\delta_{\beta}^{\sigma}+k_{12\beta}I_{12}^{\alpha\sigma}\ell_{\alpha}-\eta_{\beta\alpha}I_{12}^{\rho\alpha}\ell_{\rho}(\ell+k_{12})^{\sigma},
\end{eqnarray}
and after applying double copy for $H_{12}^{\mu\nu}$ we arrive at the result 
\begin{eqnarray}
n^{\text{GR+A}}(12,3,4;\ell) = n^{[0]}([1,2],3,4;\ell)n^{[1]}([1,2],3,4;\ell)
\end{eqnarray}
where
\begin{eqnarray}
n^{[0]}([1,2],3,4;\ell) &=& (\varepsilon_{[1,2]}\cdot\ell)(\varepsilon_{3}\cdot(\ell+k_{12}))(\varepsilon_{4}\cdot\ell)\\
n^{[1]}([1,2],3,4;\ell) &=& (\varepsilon_{[1,2]}\cdot\ell\delta_{\beta}^{\alpha}-\varepsilon_{[1,2]\beta}(\ell+k_{12})^{\alpha}+k_{12\beta}\varepsilon_{[1,2]}^{\alpha})\notag\\
&&\times(\varepsilon_{3}\cdot(\ell+k_{12})\delta_{\gamma}^{\beta}-\varepsilon_{3\gamma}(\ell+k_{123})^{\beta}+k_{3\gamma}\varepsilon_{3}^{\beta})\notag\\
&&\times(\varepsilon_{4}\cdot\ell\delta_{\alpha}^{\gamma}-\varepsilon_{4\alpha}\ell^{\gamma}+k_{4\alpha}\varepsilon_{4}^{\gamma}).
\end{eqnarray}

From the previous we can immediately write the numerator for the bubbles as 
\begin{eqnarray}
n^{\text{GR+A}}(12,34;\ell) = n^{[0]}([1,2],[3,4];\ell)n^{[1]}([1,2],[3,4];\ell).
\end{eqnarray}
With these we obtain the rest of the contributions via relabelling.

We can now built a general expression for the numerator of the $n$-gon that does not involve any contact terms in the following way
\begin{eqnarray}\label{eq:num-ngon}
n^{\text{GR+A}}(P_1,P_2,\dots,P_n;\ell) = n^{[0]}([P_1],[P_2],\dots,[P_n];\ell)n^{[1]}([P_1],[P_2],\dots,[P_n];\ell)
\end{eqnarray}
here $[P_i]$ represents the bracketed words that appear in the external currents that satisfy colour-kinematics duality. This can be a sum, if we look at \eqref{eq:DC-BG23}, for example the external-leg bubble diagram is zero after integration, but let us write an example of one those numerators 
\begin{eqnarray}\label{eq:num-ngon}
n^{\text{GR+A}}(123,4;\ell) &=& n^{[0]}([[1,2],3],4;\ell)n^{[1]}([[1,2],3],4;\ell)+n^{[0]}([[1,3],2],4;\ell)n^{[1]}([[1,3],2],4;\ell)\notag\\
&&+n^{[0]}([[2,3],1],4;\ell)n^{[1]}([[2,3],1],4;\ell),
\end{eqnarray}
each one will have its respective pole assigned.

Therefore, if we propose that the metric density coupled to a Proca field admits tree-level double copy, this property is inherited by the one loop-numerators, up to contact terms.

At this point we can to mention another advantage of this found double copy at one-loop. The external trees from the original gravitational action \eqref{eq:LL-action} have terms proportional to the flat metric $\eta^{\mu\nu}$, which lead to factors of $\ell^2$ in the integrand, responsible for many spurious divergences during the integration. Such contributions are automatically excluded in the double-copy formulation.

In the following sections we will investigate which other amplitudes can be calculated from the numerators we found by squaring each or them separately or replacing with a colour factor. 

\section{One-loop four-graviton amplitudes}

From the numerators found in the previous section we can calculate the four-graviton amplitude for three different cases by squaring the numerators. First, we will find the amplitude with a scalar loop from the square of the scalar numerator. Second, the Proca loop that we presented on the previous section. The third case will be discussed in the next section. 

\subsection{Scalar loop}

The metric density coupled to scalar field has a simple action that we write here just for reference, but no for the calculations of the amplitudes we present here. The action for the scalar coupled to the metric density reads
\begin{eqnarray}\label{eq:gra-scal}
S_{\phi}=\int d^4x\,\big(\tfrac{1}{2}\mathfrak{g}^{\mu\nu}\partial_{\mu}\phi\partial_{\nu}\phi-\tfrac{1}{2}(\mathfrak{g})^{-1/2}m^2\phi^2\big).
\end{eqnarray}
We notice that the contact terms are exclusively on the mass part, so the massless case has only a cubic interaction.

Our calculation of the four-graviton amplitudes with the scalar propagating in the loop, considering both the massive and massless cases, arises entirely from the square of $n^{[0]}$, namely
\begin{eqnarray}\label{eq:scal-dc}
n^{\text{GR+$\phi$}}(P_1,P_2,\dots,P_n;\ell) = n^{[0]}([P_1],[P_2],\dots,[P_n];\ell)n^{[0]}([P_1],[P_2],\dots,[P_n];\ell),
\end{eqnarray}
with the respective four-point conditions imposed.

In the four-point case, \eqref{eq:scal-dc} is satisfied diagram by diagram after integration for the massive and massless cases, that is if we compare the results with the ones from the diagrams that come from the action \eqref{eq:gra-scal}, since at four points there are no contributions from the contact terms sometimes even before integration. When the loop is massive, the equality stops holding for diagrams that involve non-zero contributions from the contact terms in cases with a higher number of points. For instance, in the five-graviton MHV amplitude, the relation in \eqref{eq:scal-dc} holds for nearly all contributions, with exceptions arising in six specific box permutations and two triangle permutations.

In the all-plus and mostly-plus amplitudes, \eqref{eq:scal-dc} is true for any number of points also in the massive case. 

The results from the amplitudes we obtained on this case agree with \cite{Bern:2021ppb} for the massless and massive cases. Here we present them with our conventions and only in the large mass limit to keep the expressions short, these are 
\begin{eqnarray}
\mathcal{M}_4^{\text{GR+$\phi$}}(1^+,2^+,3^+,4^+) &=& \mathcal{K}^2_1\frac{(2s+u)}{16st^3}\left(\frac{u}{504 m^2}+\frac{(s^2+su+u^2)^2}{3780st m^4}+\frac{u(s^2+su+u^2)}{7620 m^6}+\dots\right),\notag\\\\
\mathcal{M}_4^{\text{GR+$\phi$}}(1^-,2^+,3^+,4^+) &=& \mathcal{K}^2_2 t^2\left(\frac{1}{5040 m^2stu}+\frac{1}{6306300 m^8}+\dots\right),\\
\mathcal{M}_4^{\text{GR+$\phi$}}(1^-,2^-,3^+,4^+) &=& \mathcal{K}^2_3 \left(\frac{1}{6300 m^4}+\frac{s}{41580 m^6}+\frac{81s^2 + 7(su + u^2)}{15135120m^8}+\dots\right)
\end{eqnarray} 
where
\begin{eqnarray}
\mathcal{K}_1 &=& (\varepsilon_1\cdot k_3)(\varepsilon_2\cdot k_3)(\varepsilon_3\cdot k_2)(\varepsilon_4\cdot k_3),\\
\mathcal{K}_2 &=& (\varepsilon_1\cdot k_2)(\varepsilon_2\cdot k_3)(\varepsilon_3\cdot k_2)(\varepsilon_4\cdot k_2),\\
\mathcal{K}_3 &=& (\varepsilon_1\cdot k_2)(\varepsilon_2\cdot k_1)(\varepsilon_3\cdot k_2)(\varepsilon_4\cdot k_2).
\end{eqnarray}

The expressions for the full result, at any mass, in term of scalar integrals are also found in the appendix F of \cite{Bern:2021ppb}.

\subsection{Vector loop}

We go back to the original calculation with a Proca field on the loop. As we mention before, the all-plus and mostly-plus amplitudes can be calculated using double copy only. The result we reproduce for these two amplitudes is well-known for the massless and massive case, where there is a relation of direct proportionality with the amplitudes with a scalar loop that we presented on the previous section. For massive and massless vectors on the loop, we found that the expected relations hold, i.e.
\begin{eqnarray}
\mathcal{M}_4^{\text{GR+A}}(1^{\pm},2^+,3^+,4^+) &=& 3\mathcal{M}_4^{\text{GR+$\phi$}}(1^{\pm},2^+,3^+,4^+).
\end{eqnarray} 

The next case is the MHV amplitude. First we go to the massless case, that being a photon on the loop. The amplitude reads
\begin{eqnarray}
\mathcal{M}_4^{\text{GR+$\gamma$}}(1^-,2^-,3^+,4^+) &=& \frac{\mathcal{K}^2_3}{120s^8}\bigg(42(t^6 + u^6)+3(9-20\pi^2)ut(t^4+u^4)\notag\\
&&\qquad\quad -90u^2t^2(u^2+t^2)-30(5+2\pi^2)t^3u^3\notag\\
&&\qquad\quad +2\big(13(t^6-u^6)+55u^2t^2(u^2+t^2)+53ut(t^4+u^4)\big)\log\Big(\frac{u}{t}\Big)\notag\\
&&\qquad\quad -60tu(t^4 + t^2 u^2 + u^4)\log^2\Big(\frac{u}{t}\Big)\bigg),
\end{eqnarray}
which is the expected result that we checked by comparison with the amplitude from the action \eqref{eq:action-proca}, without using double copy (see also e.g. \cite{Dunbar:1994bn}).

When the loop is a Proca field the result has the same structure of the massive gluon in the loop, but with different coefficients. The explicit value of the amplitude will be on the appendix and here we present the large mass expansion 
\begin{eqnarray}\label{eq:mhv-lm}
\mathcal{M}_4^{\text{GR+A}}(1^-,2^-,3^+,4^+) = \mathcal{K}^2_3\left(\frac{31}{12600 m^4} + \frac{53s}{277200 m^6} + \frac{8072(t^2+u^2)+13113tu}{151351200 m^8}+\dots\right).\notag\\
\end{eqnarray}

Since we already studied the amplitudes with a scalar and vector in the loop, we can make a useful observation before moving to the next section. If we analyse \eqref{eq:intg-gr}, and all the expressions derived from it, we notice that the scalar (orbital) contribution can be identified already at the level of the loop current $\mathcal{J}_{P\alpha\beta}(\ell)$. In particular
\begin{eqnarray}
\mathcal{J}_{P\alpha\beta}(\ell)=\mathcal{J}_{P}^{\text{[scal]}}(\ell)\eta_{\alpha\beta}+\mathcal{J}^{\text{[spin-1]}}_{P\alpha\beta}(\ell)
\end{eqnarray}
where 
\begin{eqnarray}
\mathcal{J}_{P}^{\text{[scal]}}(\ell)=\frac{I_{P}^{\mu\nu}\ell_{\mu}\ell_{\nu}-E_Pm^2}{(\ell+k_P)^2-m^2},
\end{eqnarray}
is also the result we could also obtain from the action \eqref{eq:gra-scal}.

This decomposition is reflected at the integrand level not only for the four-point case, but even for the full $n$-point one-loop expression
\begin{eqnarray}
I_n^{\text{GR+A}}=D\,I_n^{\text{GR}+\phi}+I_n^{\text{spin}-1},
\end{eqnarray}
and, after integration, in the amplitude as 
\begin{eqnarray}
M_n^{\text{GR+A}}=M_n^{\text{orb}}+M_n^{\text{spin}-1}.
\end{eqnarray}
Here $M_n^{\text{orb}}$ is the \emph{universal orbital} (scalar-like) part and $M_n^{\text{spin-1}}$ the spin-dependent part coming from the Proca field. The important point is that this separation is manifest from the beginning, as it appears already in $\mathcal{J}_{P\alpha\beta}(\ell)$ and propagates unchanged through the integrand construction.

In the following section we will continue the exploration of amplitudes for other theories from this numerators, but the focus will be entirely on the cases where contact terms vanish.

\section{Amplitudes from single and double copy}

\subsection{Yang-Mills from single copy}

For the final case, if we consider the square of the vector numerator, this can be interpreted as corresponding to a spin-2 particle circulating in the loop. While for the scalar and vector cases we can explicitly control the contact terms that may arise, in the spin-2 case we restrict ourselves to the configurations where such terms vanish. Namely, the all-plus and mostly-plus amplitudes.

We remember that by taking the \emph{square root} of gravity, we can obtain its associated gauge theory through the so-called single-copy procedure \cite{Monteiro:2014cda}. In our setup, the double-copy factorization holds only for the terms without contact contributions, so the single copy can be consistently applied only to these same helicity configurations. This corresponds to replacing one of the kinematic numerators by a color factor $c_i$, leading to two distinct gauge-theory amplitudes. We now examine the resulting numerators for the integrands in \eqref{eq:num-ngon}.
\begin{eqnarray}\label{eq:sing-num}
n_i^{\text{YM+$\phi$}}(\ell) &=& c_in_i^{[0]}(\ell),\\
n_i^{\text{YM+A}}(\ell) &=& c_in_i^{[1]}(\ell).
\end{eqnarray}

Both numerators in \eqref{eq:sing-num} are for external gluons, what changes is the particle in the loop. On the first line we have and scalar on the second a vector (both massive or massless), but there are some subtleties to take into account here. Our starting theory is completely abelian, this means that first we have to promote the scalar field to take values in a representation of the non-abelian gauge group where the external gluons live, we will take the adjoint representation here. for the second case we would have a coupling of Yang-Mills with a Proca field or a photon, such a coupling does not exist so a solution is again to promote the vector field on the loop to live on the adjoint representation.  

The promotion of the particles on the loop to live on a certain representation of the gauge group for Yang-Mills is performed by the choice of the colour factors in \eqref{eq:sing-num}. The kinematic numerator for the scalar loop is the same as the one obtained by Feynman rules from the action for Yang-Mills-scalar, but the coupling with a vector field on the loop seems different from what comes from pure Yang-Mills. Let us see what the amplitudes tell us. 

The result for one-loop four-gluon amplitude is very well-known since many years (see e.g. \cite{Bern:1992ad} for review), and we will not reproduce it here. The important part to remark here is that if we use the numerators from single copy \eqref{eq:sing-num}, the relation between the gluon and scalar amplitudes with such numerators is the following
\begin{eqnarray}
\mathcal{M}_4^{[1]}(1^{\pm},2^+,3^+,4^+) &=& 3\mathcal{M}_4^{[0]}(1^{\pm},2^+,3^+,4^+),
\end{eqnarray} 
which does not reproduce the correct proportionality constant between the amplitudes for pure Yang-Mills and Yang-Mills-scalar for these polarizations. What is happening here is that once we promoted the abelian vectors on the loop to gluons our method of integrations does not give us the correct degrees of freedom and we would need an additional contribution from a ghost loop.

The combination that give us the correct number of degrees of freedom for pure Yang-Mills is 
\begin{eqnarray}
n_i^{\text{YM}}(\ell) = c_i(n_i^{[1]}(\ell)-n_i^{[0]}(\ell)),
\end{eqnarray}
being consistent with the ghost for Yang-Mills behaving like a scalar in the loop. We return to our example we see that the partial amplitude 
\begin{eqnarray}
\mathcal{M}_4^{\text{YM}}(1^{\pm},2^+,3^+,4^+) &=& 2\mathcal{M}_4^{[0]}(1^{\pm},2^+,3^+,4^+),
\end{eqnarray}
now has the correct constant of proportionality. 

\subsection{Gravity from double copy}

The turn now is to apply double copy in order to obtain pure gravity amplitudes. If we naively take the square of the vector numerator $n_i^{[1]}(\ell)\times n_i^{[1]}(\ell)$, we could interpret this as an amplitude of external gravitons with a massive spin-2 particle on the loop, but from the lesson we learned in the case of Yang-Mills, this will not have the correct degrees of freedom of a graviton in the massless case. We can check that directly here  
\begin{eqnarray}
\mathcal{M}_4^{[1]\times[1]}(1^{\pm},2^+,3^+,4^+) &=& 9\mathcal{M}_4^{\text{GR+$\phi$}}(1^{\pm},2^+,3^+,4^+).
\end{eqnarray}

Now, if we take the square of numerators that give the correct degrees of freedom for Yang-Mills $(n_i^{[1]}(\ell)-n_i^{[0]}(\ell))\times(n_i^{[1]}(\ell)-n_i^{[0]}(\ell))$, the amplitude would be $4\mathcal{M}_4^{\text{GR+$\phi$}}(1^{\pm},2^+,3^+,4^+)$. Therefore, the correct numerators for gravity, also for the massive loop are
\begin{eqnarray}\label{eq:gr-adc}
n_i^{\text{GR}}(\ell)=n_i^{[1]}(\ell)\times n_i^{[1]}(\ell)-2n_i^{[0]}(\ell)\times(n_i^{[1]}(\ell)-n_i^{[0]}(\ell)),
\end{eqnarray}
Again being consistent with the ghost for gravity behaving like a vector on the loop. Back to the four-graviton amplitude
\begin{eqnarray}\label{eq:gr-amps}
\mathcal{M}_4^{\text{GR}}(1^{\pm},2^+,3^+,4^+) &=& 5\mathcal{M}_4^{\text{GR+$\phi$}}(1^{\pm},2^+,3^+,4^+),
\end{eqnarray}
the numerators \eqref{eq:gr-adc} reproduce the correct result.

We see that if we star from the numerators from our abelian theory, in order to obtain pure gravity, the relation \eqref{eq:gr-adc} is not exactly a double copy. The main reason for that is that our numerators do not satisfy colour-kinematics duality, and even simple amplitudes such as \eqref{eq:gr-amps} can be obtained from double copy from self-dual Yang-Mills, but this theory exhibits colour-kinematic duality even at the off-shell level \cite{Monteiro:2011pc,Boels:2013bi,Correa:2024mub}.     

\section{Conclusions}

In this work we analysed one-loop four-graviton amplitudes using the Landau-Lifshitz tensor density formulation of gravity, coupled to a Proca field circulating in the loop. We found that the integrands without contact term exhibit a manifest double-copy factorization in the numerators. Using the same abelian numerators, we calculated a broader class of one-loop four-graviton and four-gluon amplitudes with massless and massive particles in the loop -- including scalar, vector, and tensor contributions. 

Remarkably, the double- and single-copy structures arise even though the one-loop numerators do not satisfy color-kinematics duality, this condition was imposed only for the external trees attached to the loop. 

There are three works in progress following the present one: The first one is related to the analysis of the contact terms to calculate amplitudes with more configurations in their polarizations than the all-plus and the mostly-plus ones. The second one requires modern integration techniques, like integrations by parts \cite{Chetyrkin:1981qh,Laporta:2000dsw} and differential equations methods \cite{Kotikov:1990kg,Henn:2013pwa}, to go to a higher number of external gravitons, starting with the five-point case. The third case of present study is the calculation of one-loop amplitudes with external massive vectors to analyse the corrections to the gravitational potential on this setting.  

Future directions include extending this framework to higher loop orders, as well as exploring phenomenological applications. In particular, the interaction of gravity with massive scalar and abelian vector fields could provide valuable insight into models of the dark sector where such fields play a central role \cite{Jaeckel:2016jlh}.

\section*{Acknowledgements}

We thank Humberto Gomez, Renann Lipinski, Alexander Quintero Velez, and Christian Schubert for valuable insights and discussions. Special thanks also to Andres Felipe Rendon for hospitality during some steps of this project. This work was supported by the Czech Academy of Sciences under the Project No. LQ100102101.

\appendix

\section{Explicit expression for the MHV amplitude}

In this appendix we will present the full result for the MHV amplitude that we presented in \eqref{eq:mhv-lm} in the large mass limit. Our result will be written in terms of the finite part of the scalar integrals that come from the PV tensor reduction
\begin{eqnarray}
B_0(u,m) &=& \frac{\sqrt{u \left(u-4 m^2\right)}}{u} \log \Bigg(\frac{\sqrt{u \left(u-4 m^2\right)}+2 m^2-u}{2 m^2}\Bigg),\\
C_0(t,m) &=& \frac{1}{2t}\log^2\Bigg(\frac{\sqrt{t \left(t-4 m^2\right)}+2 m^2-t}{2 m^2}\Bigg),\\
D_0(s,t,m) &=& I_4^{\text{fin}}(s,t,m),
\end{eqnarray}
where $I_4^{\text{fin}}(s,t,m)$ is the finite part of the massive scalar box integral
\begin{eqnarray}
I_4(s,t,m) = \int \frac{d^D\ell}{i\pi^{D/2}}\frac{1}{(\ell^2-m^2)((\ell+k_1)^2-m^2)((\ell+k_{12})^2-m^2)((\ell-k_4)^2-m^2)},
\end{eqnarray}
which results comes in terms of $\log$ and $\text{Li}_2$ functions. The other integrals come from permutations on the external legs.

With these definitions the result for the amplitude reads
\begin{eqnarray}
\mathcal{M}_4^{\text{GR+A}}(1^-,2^-,3^+,4^+) &=& \mathcal{K}_3^2 f(s,t,u)
\end{eqnarray}
where
\begin{eqnarray}
f(s,t,u) &=& \frac{m^2 \left(9 m^2 \left(s^2+u^2\right)-8 s t u\right) \left(\frac{1}{\epsilon }+\log\left(\frac{\mu^2}{m^2}\right)\right)}{6 s^2 t^2 u^2}\notag\\
&& +\frac{1}{120s^6t^2u^2}\bigg(6 m^4 s^2 (18 u^4-7t^4-60t^3u+129t^2u^2-10tu^3)\notag\\
&&\qquad\qquad\quad +2m^2stu(44t^4-185t^3u+22t^2u^2-185tu^3+44u^4)\notag\\
&&\qquad\qquad\quad +3t^2u^2(14t^4-19t^3u-6t^2u^2-19tu^3+14u^4)\bigg)\notag\\
&&\ +\frac{1}{60s^7t^2}\bigg(6m^4s^2(8u^3-73t^3+147t^2u+48tu^2)\notag\\
&&\qquad\qquad\quad +2m^2st(133t^4-154t^3u-45t^2u^2-236tu^3-28u^4)\notag\\
&&\qquad\qquad\quad -t^2(t-u)(13t^4-53t^3u-42t^2u^2-53tu^3+13u^4)\bigg)B^0(u,m)\notag\\
&& +\frac{1}{60s^7u^2}\bigg(6m^4s^2(8t^3+48t^2u+147tu^2-73u^3)\notag\\
&&\qquad\qquad\quad +2m^2su(133u^4-28t^4-236t^3u-45t^2u^2-154tu^3)\notag\\
&&\qquad\qquad\quad +u^2(t-u)(13t^4-53t^3u-42t^2u^2-53tu^3+13 u^4)\bigg)B^0(u,m)\notag\\
&& -\frac{1}{s^8}\bigg(12m^6s^3-6m^4s^2(2t^2-tu+2u^2)+2m^2s(t^4-4t^3u-t^2u^2-4tu^3+u^4)\notag\\
&&\qquad\qquad\quad +tu(t^4+t^2u^2+u^4)\bigg)\big(tC^0(t,m)+uC^0(u,m)\big)\notag\\
&& +\frac{tu}{2s^8}\bigg(48m^6s^3-12m^4s^2(3t^2+tu+3u^2)+4m^2s(t^4-2t^3u-2tu^3+u^4)\notag\\
&&\qquad\qquad\quad +tu(t^4+t^2u^2+u^4)\bigg)D^0(t,u,m)\notag\\
&& +\frac{m^2}{s^4}\big(3m^4-4m^2s+s^2\big)\big(D^0(s,t,m)+D^0(s,u,m)+D^0(t,u,m)\big)\notag\\
\end{eqnarray}

\providecommand{\href}[2]{#2}\begingroup\raggedright\endgroup
\end{document}